\documentclass[preprint2,twoside]{hwo}

\usepackage{lipsum}

\newcommand{\rpro}{\mbox{{\it r}-process}}

\input{hwo.h}

\begin{document}

\title{\textbf{\LARGE Habitable Worlds Observatory:\ The Nature of the Astrophysical r-process}}

\author {\textbf{\large Ian U.\ Roederer,$^{1}$ Rana Ezzeddine,$^2$ }}

\affil{$^1$\small\it Department of Physics, North Carolina State University, Raleigh, North Carolina, USA; \email{iuroederer@ncsu.edu}}

\affil{$^2$\small\it Department of Astronomy, University of Florida, Gainesville, Florida, USA}


\author{\footnotesize{\bf Endorsed by:}
Shiva Agarwal (Western Michigan University),
Narsireddy Anugu (Georgia State University),
St\'{e}phane Blondin (Aix Marseille Univ/CNRS/LAM),
Daniel Brethauer (UC Berkeley),
Mark Elowitz (Network for Life Detection (NfoLD)),
Vincent Esposito (Chapman University),
Luca Fossati (Space Research Institute, Austrian Academy of Sciences),
Emma Friedman (NASA GSFC),
Natalie Hinkel (Louisiana State University),
Pierre Kervella (Paris Observatory \& CNRS IRL FCLA),
Joris Josiek (ZAH/ARI, Universit\"{a}t Heidelberg),
Eunjeong Lee (EisKosmos (CROASAEN), Inc.),
Pranav Nalamwar (University of Notre Dame),
Gijs Nelemans (Radboud University),
Mardav Panwar (Meerut, College, Meerut),
Julia Roman-Duval (Space Telescope Science Institute),
Blair Russell (Chapman University),
Farid Salama (NASA Ames Research Center),
Melinda Soares-Furtado (UW-Madison),
Shivani Shah (North Carolina State University),
Josh Simon (Carnegie Observatories)
}



\begin{abstract}
We present the science case for characterizing the origin of the 
heaviest elements on the periodic table, with a focus on those produced by the
rapid neutron-capture process (\rpro),
using the Habitable Worlds Observatory (HWO).~
High-resolution ultraviolet (UV) spectroscopy can increase
the number of \rpro\ elements detectable in
cool stars by more than 50\% relative to optical and infrared spectra.
These elements are key to characterizing the physical conditions
that govern the \rpro\ and identify the 
nature, sites, and environments where \rpro\ events occurred.
HWO has the potential to greatly expand the sample
of stars where rarely studied heavy elements can be detected
beyond the Solar neighborhood to the 
Galactic halo, globular clusters, and dwarf galaxies.
  \\
  \\
\end{abstract}

\vspace{2cm}

\section{Science Goal}

What is the origin of the elements all across the periodic table?
The heaviest elements listed on the periodic table, 
including silver, gold, platinum, and uranium, 
may be formed through a process of rapidly adding neutrons to 
lighter atomic nuclei.  
This rapid neutron-capture process, or \rpro\, 
is one of the fundamental ways that stars produce the heaviest elements.
These elements can be detected in stars that formed 
after \rpro\ nucleosynthesis events.  
We can learn about these events by studying the patterns, 
or amounts, of the \rpro\ elements found in ancient stars.  
These patterns reveal information about the nature, sites, 
and environments where \rpro\ events occurred.
High-resolution ultraviolet (UV) 
(1700 $< \lambda <$ 3100~\AA) spectroscopy 
is critical to collect the data we need to make these measurements, 
better understand these fascinating events, 
and learn about the origins of the elements in the world around us.

Understanding how these elements are made 
is a key part of addressing two of the broad themes 
identified by the Astro 2020 Decadal Survey \citep{astro2020}: 
``New Messengers and New Physics”'' and ``Cosmic Ecosystems.'' 
Only HWO has the high spectral resolution in the UV 
necessary to detect these elements.

\section{Science Objective}

Collecting high-resolution UV spectra enables a 50\% advance, 
relative to optical and infrared spectra, 
in the number of \rpro\ elements that can be detected in cool
(FGK) stars.

The science objective is to generate a chemical inventory 
of which elements are present, and in what amounts, 
for as many \rpro\ elements in as many stars as possible. 
The resulting abundance patterns---including the 
element-to-element dispersion (or lack thereof) and overall amounts---can be 
compared with model predictions to reveal the physical conditions, 
sites, and environments where the \rpro\ occurs.

The Milky Way and Local Group of dwarf galaxies contain metal-poor stars
that are enhanced in \rpro\ elements.
More than 200 \rpro-enhanced stars are known at present.
Ongoing and future large, targeted spectroscopic surveys,
such as the Galactic Archaeology with HERMES survey
(GALAH; e.g., \citealt{matsuno21}),
$R$-Process Alliance (RPA; e.g., \citealt{hansen18})
and the 4-meter Multi-Object Spectroscopic Telescope project
(4MOST; e.g., \citealt{skuladottir23})
are increasing the number of \rpro-enhanced stars
known in the Milky Way and Local Group.
The hundreds or thousands of \rpro-enhanced 
stars discovered in these surveys have
the potential to significantly transform our understanding 
of the origins and sites of the \rpro.

\begin{figure*}[ht]
\begin{center}
\includegraphics[width=0.8\textwidth]{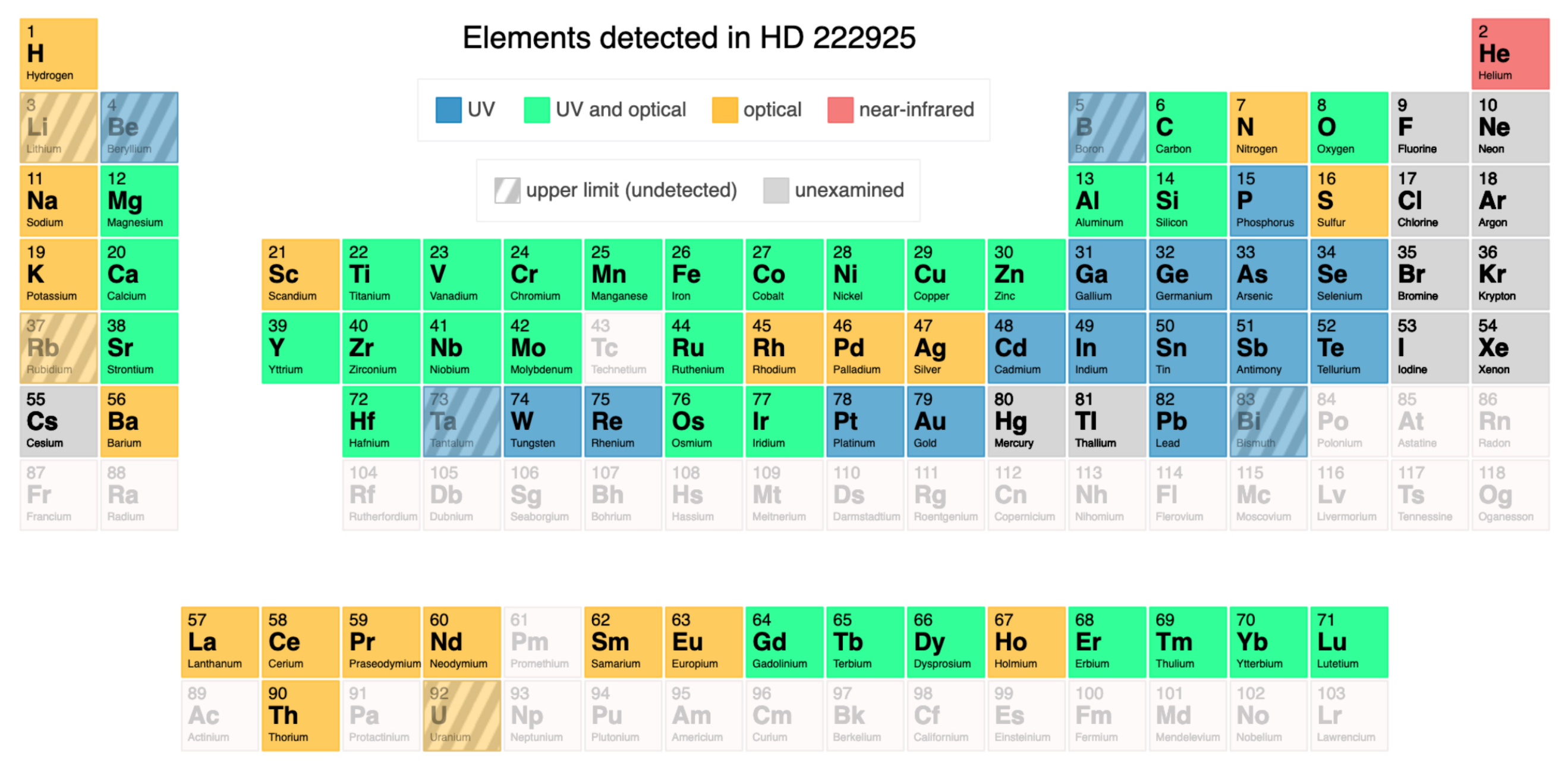}
\caption{\small 
Periodic table showing the elements examined 
in the best-studied \rpro-enhanced star at present, \mbox{HD~222925}, 
highlighting the wavelength regions where various elements 
can be detected \citep{roederer22a}.
Many other \rpro-enhanced stars are known, 
but only this one is bright enough 
to collect high-quality UV spectra from 2000~\AA\ redward with STIS.~
\label{fig:periodictable}
}
\end{center}
\end{figure*}

The UV and optical spectra of these stars are rich 
with transitions of more than 40 r-process elements 
\citep[Figure~\ref{fig:periodictable};][]{roederer22a}. 
Each of these stars likely reflects the yields of a single \rpro\ event, 
such as a neutron-star merger or exotic kind of supernova. 
The detailed abundance patterns can be modeled 
to reveal the nature of each site and constrain the physics at the 
\rpro\ site \citep{ji18th,yong21nature,holmbeck23}.
Optical spectra are useful to identify and characterize 
some aspects of the \rpro\ abundance patterns.
UV spectra are required to detect elements at the three \rpro\ peaks 
(Se, Te, Pt; \citealt{denhartog05,roederer12b,roederer12a}),
which are among the most sensitive to the physics at the \rpro\ site
\citep{eichler15,mumpower16}. 
Other elements that are only detectable in the UV 
reveal the role of fission of transuranic elements in the r-process
\citep{roederer23b}. 
$R$-process-enhanced stars exhibit abundance variations 
in the lightest and heaviest \rpro\ elements, 
and larger samples of stars with more complete chemical inventories
are necessary to characterize these variations 
and assess how frequently they occur \citep{barklem05heres,holmbeck20}.

The observational challenge is that there is only one source 
of high-resolution UV spectra, 
the Space Telescope Imaging Spectrograph (STIS) 
on the Hubble Space Telescope (HST).~ 
Although more than 200 \rpro-enhanced stars are currently known, 
only $\approx$~10 such stars are bright enough 
to have been observed over the last quarter-century with STIS.~
Only one of those stars, \mbox{HD~222925},
is bright enough to collect the high-quality spectra 
($R \equiv \lambda/\Delta\lambda >$ 100,000 and $20 < {\rm S/N} < 40$ 
across 2000--3100~\AA) 
required for the detection of all 42 \rpro\ elements shown in 
Figure~\ref{fig:periodictable}.
In contrast, according to data in the Hypatia Catalog 
\citep{hinkel14}, the heavy elements 
Sr, Y, Zr, Ba, La, Nd, and Eu that are readily detectable in optical spectra 
of cool stars have been studied in $\approx$~3000--6000 stars.

\begin{figure}[ht]
\begin{center}
\includegraphics[width=0.45\textwidth]{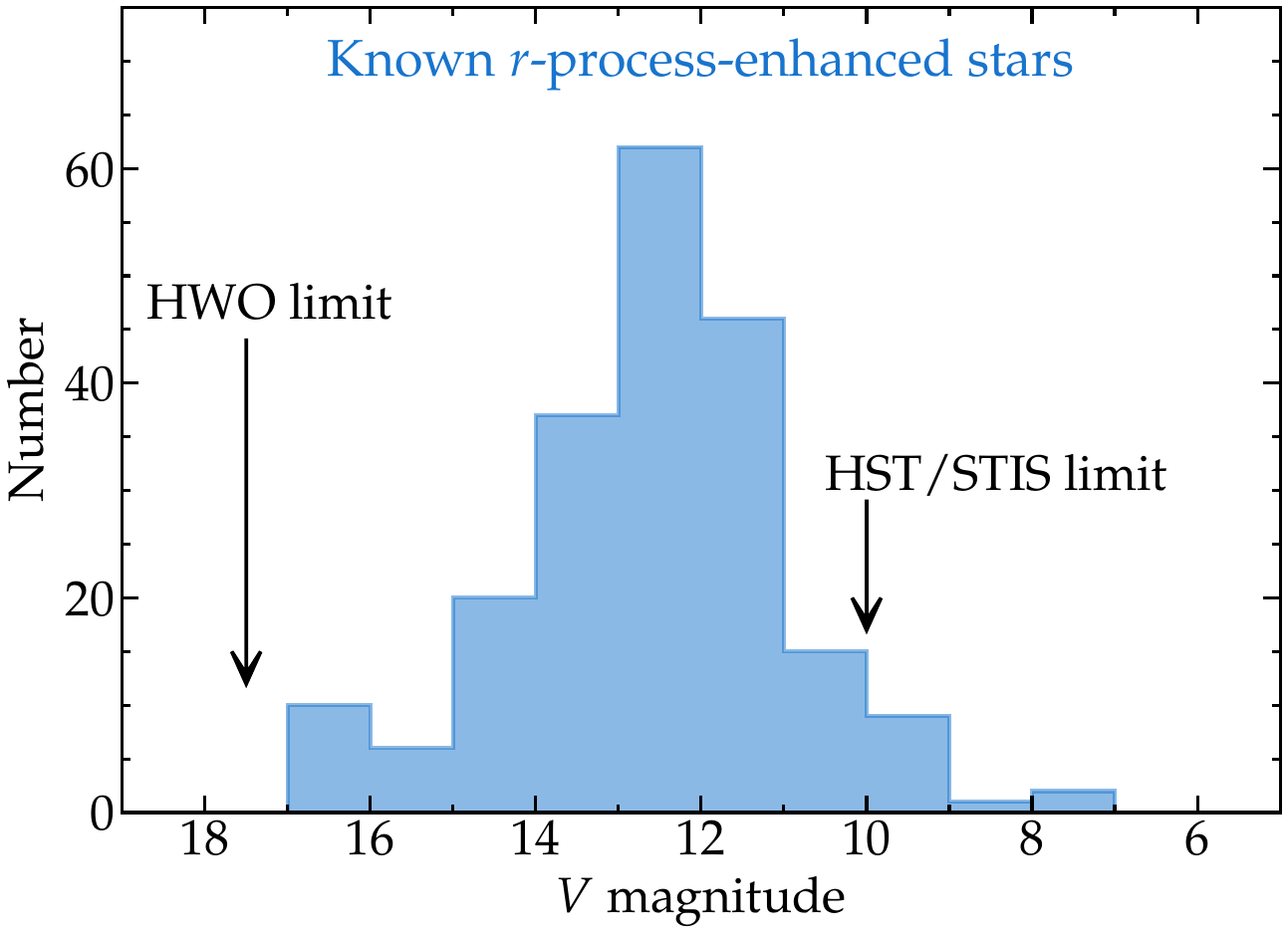}
\caption{\small 
Histogram of the magnitudes of known \rpro-enhanced stars, 
with limits for HST ($V \sim$~10) and HWO ($V \sim$~17.5).
HWO would enable high-resolution spectroscopic observations 
for virtually any metal-poor star observable today
with high-resolution optical spectroscopy. 
\label{fig:vmag}
}
\end{center}
\end{figure}

The HWO has the potential to expand the stellar sample dramatically, 
such that virtually any star accessible today 
to high-resolution spectrographs on ground-based optical telescopes
could be observable in the UV (Figure~\ref{fig:vmag}).
Success entails observing $\sim$~100 UV lines
of about 20 elements produced by the \rpro\ 
in $\gg$~10~stars across a range of Galactic environments.

\section{Physical Parameters}

Atomic physics dictates that the ionization states 
commonly found in the atmospheres of metal-poor stars 
on the main sequence, subgiant branch, 
red giant branch, and horizontal branch
(effective temperatures in the $\approx$~4000--6500~K range)
are neutral and singly ionized.
The strongest transitions of these ionization states 
of some \rpro\ elements are in the optical (3100--10,000~\AA), 
but for about 20 \rpro\ elements 
(Figure~\ref{fig:periodictable}) 
they are in the UV in the $\sim$~1700--3100~\AA\ range
(Table~1).
Many of these elements are listed toward the right side of the periodic table, 
which are the ones located at or near the so-called \rpro\ peaks, 
where the chemical abundances are most sensitive 
to the physical conditions of the \rpro\ events.
Detecting absorption lines of these elements 
enables us to form a more complete template 
of the element-to-element chemical abundance patterns 
that are created by \rpro\ events.
This template pattern, or patterns, 
constrains the physics at the \rpro\ site or sites.

\begin{table}[!ht]
\caption{Wavelengths of Key Lines in the UV}
\smallskip
\begin{center}
{\small
\begin{tabular}{cc}  
\tableline
\noalign{\smallskip}
\textbf{Species} & \textbf{Wavelength (\AA)} \\
\noalign{\smallskip}
\tableline
\noalign{\smallskip}
Ge~\textsc{i}  &  3039 \\
Ta~\textsc{ii} &  2752 \\
Au~\textsc{i}  &  2675 \\
Lu~\textsc{ii} &  2615 \\
Cd~\textsc{i}  &  2288 \\
Os~\textsc{ii} &  2282 \\
Te~\textsc{i}  &  2259 \\
Ir~\textsc{ii} &  2255 \\
Re~\textsc{ii} &  2214 \\
Pb~\textsc{ii} &  2203 \\
Sb~\textsc{i}  &  2185 \\
W~\textsc{ii}  &  2118 \\
Ga~\textsc{i}  &  2090 \\
Se~\textsc{i}  &  2074 \\
Pt~\textsc{ii} &  2057 \\
Mo~\textsc{ii} &  2020 \\
Hg~\textsc{ii} &  1942 \\
Tl~\textsc{ii} &  1908 \\
Sn~\textsc{ii} &  1899 \\
As~\textsc{i}  &  1890 \\
Ru~\textsc{ii} &  1875 \\
I~\textsc{i}   &  1830 \\
Bi~\textsc{ii} &  1791 \\
Sb~\textsc{ii} &  1606 \\
Br~\textsc{i}  &  1540 \\
In~\textsc{ii} &  1586 \\
\noalign{\smallskip}
\tableline\
\end{tabular}
}
\end{center}
\end{table}
\noindent

$R$-process-enhanced stars are found throughout the Milky Way
and its neighboring satellite stellar systems.
Only a handful of these stars are nearby enough to observe with HST.~
Each star retains the \rpro\ elements produced in the \rpro\ event(s)
that preceded it, but in the case of \rpro-enhanced stars the 
\rpro\ element pattern is dominated by a single event.
The star-to-star uniformity or variation among these \rpro\
element abundance patterns constrains the range of physical parameters 
and/or sites, such as neutron-star mergers or exotic classes of supernovae, 
where \rpro\ nucleosynthesis occurs.
Detection of $\approx$~20 \rpro\ elements per star in the UV,
and $\approx$~40 \rpro\ elements overall when combined with optical spectra,
for each of 20+ stars 
in different Galactic environments
constrains the range of physical parameters and \rpro\ sites
in the early Universe.

Abundances would be derived to a precision of $<$~0.3~dex 
to confirm the basic \rpro\ pattern for a previously undetected element, 
and $<$~0.1~dex to constrain nuclear physics parameters 
and characteristics of various \rpro\ sites.
Existing atomic data (transition probabilities, hyperfine splitting structure,
etc.)\ are generally sufficient for this application
(e.g., \citealt{roederer12b,kramida24}).

The continuous opacity can be reliably modeled to wavelengths 
at least as short as $\approx$~2000~\AA, 
and probably even $\approx$~1800~\AA, in metal-poor stars. 
No attempts have been made to observe metal-poor stars
at wavelengths as short as $\approx$~1500~\AA, 
where metal opacities become comparable to the H$^{-}$ opacity.
This is an area that requires theoretical exploration.

High-resolution $R \sim$~100,000 UV spectroscopy
in the wavelength range of 2000--3100~\AA\ has proven successful 
for one star with $V$ = 9.0 with STIS \citep{roederer22a}.
Going to wavelengths in the range of 1500--2000~\AA\ 
is expected to open up new discovery space \citep{peterson20}, 
because it would enable detection of lines of several key \rpro\ elements 
(Br, I, Hg, Tl) that have not been studied previously.
It would also enable detection of the dominant ionization states 
of other \rpro\ elements that may be
stronger and less sensitive to uncertainties 
in the radiative transfer calculations. 

The objective is to develop a set of \rpro\ ``template'' patterns 
to constrain model predictions. 
Early studies \citep{johnson02rpro,sneden09,siqueiramello14} of $\approx$~5 
\rpro-enhanced stars demonstrated remarkable abundance uniformity 
among some element groups, but not others.
As sample sizes grew to $\approx$~10--15 stars 
\citep{barklem05heres,mashonkina14baeu,roederer14e},
infrequent \rpro\ abundance phenomena and correlations 
(or lack thereof) with other element abundance signatures 
could begin to be assessed.
As sample sizes increased further, to $\sim$~100 or more
\citep{roederer18d,ji19,holmbeck20,gudin21}, 
statistical analyses of occurrence frequencies, 
environmental dependence, and comparisons with kilonova yields 
could also be assessed.
These studies, all based on analyses of \rpro\ elements in optical spectra, 
guide our judgments about how well the present science objectives
can be achieved for \rpro\ model properties
that are best constrained by elements detectable only in UV spectra
(Table~2).

\section{Description of Observations}

The goal is to obtain high-S/N, high-resolution UV spectra 
for each \rpro-enhanced star 
to produce detections of lines of \rpro elements (Table~3).
These lines will often be blended with other stronger lines, 
so high resolution and S/N are needed. 
Experience suggests that $R$ = 40,000 is a minimum for blended lines, 
and $R \sim$~100,000 is ideal and fully sufficient 
to resolve the line profiles \citep{roederer22a}.
($R \gtrsim$ 100,000 does not add additional benefits for this science case). 
Experience suggests S/N = 20 per pixel is a minimum, while
S/N = 40--50 per pixel is sufficient.

Wavelengths as long as 3100~\AA\ should be obtainable 
to provide continuous spectral coverage with ground-based optical spectra.
Coverage to at least 2000~\AA\ is highly desirable.
Coverage to $\approx$~1500~\AA\ would open entirely new discovery space.
Spectra spanning the wavelength range should be obtainable
in as few setups as possible to maximize telescope efficiency,
because the full spectral range will be desirable for this science case.

The field-of-view (FOV) and angular resolution 
are not constraints that will drive this science case.
The FOV only needs to be large enough to observe one point-source target star
at a time on a narrow slit. 
Target stars are unlikely to be in crowded fields 
(within $\approx$~0\farcs1 or so) 
with other objects of comparable brightness 
(within $\approx$~5~mag or so). 
At 2000~\AA, the diffraction limit (1.22$\lambda/D$) 
for a 6.5~m telescope is approx 0\farcs008, 
far smaller than angular separation to any potential neighbor objects.

Target stars will likely be spread across the entire sky, 
and in few if any cases will they be within a few arcseconds 
of each other, so the number of fields is equivalent to the number of targets.

HWO is necessary because HST/STIS is not sensitive enough 
at these wavelengths for stars with $V \sim$~10 ($NUV \sim$~13--16). 
The numbers in Table~3 are based on prior experience with STIS spectra.

{\bf Acknowledgements.}
We acknowledge funding from the US National Science Foundation 
(AST~2205847 to IUR, AST-2206263 to RE)
and numerous grants 
(including GO-15657 and GO-15951 to IUR and RE)
provided by NASA through the Space Telescope
Science Institute, which is operated by the Association of 
Universities for Research in Astronomy, Incorporated, under NASA
contract NAS5-26555.

\begin{table*}[!ht]
\caption{Physical Parameters}
\smallskip
\begin{center}
{\small
\begin{tabular}{ccccc}  
\tableline
\noalign{\smallskip}
\textbf{Physical Parameter} &
\textbf{State of the Art} &
\textbf{Incremental Progress} &
\textbf{Substantial Progress} &
\textbf{Major Progress} \\
 & & \textbf{(Enhancing)} & \textbf{(Enabling)} & \textbf{(Breakthrough)} \\
\noalign{\smallskip}
\tableline
\noalign{\smallskip}
Number of elements, $N$, detected & 42 & 45 & 45 & 45 \\
in the \rpro\ pattern             & & & & \\
\noalign{\smallskip}
Number of stars where $N$ & 1 & 10 & 20 & 100 \\
elements can be detected & & & & \\
\noalign{\smallskip}
\tableline\
\end{tabular}
}
\end{center}
\end{table*}

\begin{table*}[!ht]
\caption{Observational Requirements}
\smallskip
\begin{center}
{\small
\begin{tabular}{ccccc}  
\tableline
\noalign{\smallskip}
\textbf{Observational Requirement} & 
\textbf{State of the Art} & 
\textbf{Incremental Progress} & 
\textbf{Substantial Progress} & 
\textbf{Major Progress} \\
 & & \textbf{(Enhancing)} & \textbf{(Enabling)} & \textbf{(Breakthrough)} \\
\noalign{\smallskip}
\tableline
\noalign{\smallskip}
$V$ magnitude of FGK star  & 10 & 12 & 14 & 17 \\
\noalign{\smallskip}
Wavelength range (\AA) & 2000--3100 & 2000--3100 & 1700--3100 & 1500--3100 \\
\noalign{\smallskip}
S/N after co-adds & 40 & 40 & 40 & 40 \\
\noalign{\smallskip}
Spectroscopic resolving power & 100,000 & 40,000 & 100,000 & 100,000 \\
\noalign{\smallskip}
FOV & (one point source) & (one point source) & (one point source) & (one point source) \\
\noalign{\smallskip}
Angular resolution at 2000~\AA\ & $<$ 0\farcs1 & $<$ 0\farcs1 & $<$ 0\farcs1 & $<$ 0\farcs1 \\
\noalign{\smallskip}
Number of fields & 1 & 10 & 20 & 100 \\
\noalign{\smallskip}
\tableline\
\end{tabular}
}
\end{center}
\end{table*}
\noindent

\bibliography{roederer_rprocess}

\end{document}